\documentclass[useAMS,usenatbib]{mn2e}
\usepackage{graphicx}
\usepackage{natbib}
\usepackage{color}
\newcommand{\feh}{\mbox{[Fe/H]}}

\newcommand{\mgh}{\mbox{[Mg/H]}}

\newcommand{\sih}{\mbox{[Si/H]}}
\newcommand{\cah}{\mbox{[Ca/H]}}
\newcommand{\crh}{\mbox{[Cr/H]}}
\newcommand{\znh}{\mbox{[Zn/H]}}
\newcommand{\afe}{\mbox{[$\rm \alpha$/Fe]}}

\title[Investigation of the Prompt SNe Ia progenitor nature]
{Investigation of the Prompt SNe Ia progenitor nature through the analysis of the chemical composition of globular clusters and circumgalactic clouds}

\author[I.A. Acharova,  M.E. Sharina, and E.A. Kazakov]{I.A. Acharova$\rm^{1}$\thanks{E-mail:
iaacharova@sfedu.ru (AIA)}, M.E. Sharina$\rm^{2}$\thanks{E-mail: sme@sao.ru (SME)},E.A. Kazakov$\rm^{1}$\\
$\rm^{1}$Department of Physics, Southern Federal University, 5 Zorge, Rostov-on-Don, 344090 Russia\\
$\rm^{2}$Special Astrophysical Observatory, Russian Academy of Sciences, Nizhny Arkhyz, 369167 Russia}

\begin{document}

\date{Accepted  Received }

\pagerange{\pageref{firstpage}--\pageref{lastpage}} \pubyear{2019}

\maketitle

\label{firstpage}

\begin{abstract}
A method is proposed for determining the properties of type Ia supernovae from short-lived precursors -- Prompt SNIa. 
This method is based on the assumption that this subtype of type Ia supernovae exploded into low-metallicity globular 
clusters (GCs), and is responsible for the enrichment of the high-metallicity subgroup of GCs and circumgalactic 
clouds (CGCs) with the iron peak elements. We justify that CGCs are the formation places of GCs of both subgroups. 
The accuracy of the method depends, first, on the number of GCs, the spectra of which have been studied in detail; 
second, on the number of chemical elements, the abundances of which have been worked out. 
Only those elements are of interest for this method that are produced in supernova explosions and are not 
produced at the previous stage of the stellar evolution. Our estimates of nucleosynthesis in low-metallicity 
supernova GCs are in the best agreement with the following Prompt SNIa model: Single Degenerate Pure Deflagration 
Models of white dwarfs (WDs) burning with masses in the range from $\rm 1.30  M_{\sun}$ to $1.31  M_{\sun}$ if 
carbon explodes in the centre of a WD with a low central density from 
$\rm 0.5 \cdot10^9 g/cm^3$ to $\rm 10^9 g/cm^3$.

\end{abstract}

\begin {keywords}

(Galaxy:) globular clusters: general -- (stars:) supernovae: general
-- galaxies: evolution -- galaxies: formation -- (galaxies:) quasars:
absorption lines

\end{keywords}

\section[]{Introduction}

   The possibility to consider circumgalactic clouds (CGCs) as remnants of the parent clouds, in which  globular clusters (GCs) 
have been formed, was first established by \citet{ASh18} based on the following facts. 
First, both types of objects show a bimodal distribution 
with the minimum near $\rm [Mg/H]=-0.9$. The mean values and standard deviations of the Mg abundance in GCs and CGCs with 
$\rm [Mg/H]< -0.9$  and $\rm [Mg/H]> -0.9$ are very similar. 
This coincidence implies that the clouds and GCs can be  evolutionary related. 
Second, high-metallicity CGCs are observed at redshifts smaller than $\rm z\sim 2.5$.  
At the redshifts $\rm 2.5~<z<~4$, the metallicity distribution of clouds is presented mainly by the  metal-poor component 
\citep{Raf12, Lehner16}. High-metallicity CGCs appeared for the first time at $\rm z\sim 4$ \citep{Raf12, Berg15,Jorgenson13}.
This means that high-metallicity gas has originated in the formation epoch of GCs. 
Therefore, heavier-element enrichment of the fraction of metal-poor gas has taken place 
through thermonuclear fusion products of supernovae (SNe) of the first GC  generation entering into it. 
Subsequently, the second generation has been formed from this enriched gas. 
In addition to the closeness of the metallicity distributions and the presence of GCs and CGCs at $z<2$, 
we can offer more observations in support of the hypothesis of formation of GCs from CGCs. 
The observation facts and theoretical modelling results indicate that the mass of CGCs is 
proportional to the mass of the halos of the galaxies, in the vicinity of which they are observed \citep{Perez18}. 
On the other hand, according to \citet{Blakeslee99}, the GC formation rate is universal and is $0.5-1$ GC 
per $\rm 10^9 M_{\sun}$ of a parent galaxy.  
In other words, the number of both objects we considered increases monotonically with host galaxy mass.
To justify the formulation of the problem solved in our paper, it is important to note the following. 
There is no clear relationship between \afe\ \footnote{Different authors
considered different $\rm \alpha$-process elements to calculate [$\rm \alpha$/Fe]. \citet{carney96} used 
Si, Ca, and Ti. \citet{nissen07} considered S. \citet{d16} used  Mg, Ca, and Ti.} 
and the age of GCs. 
This issue was discussed in the papers by \citet{carney96}. The decrease of \afe\ with  the increase of metallicity 
in the Galactic disc stars or in dwarf galaxies with extended star-formation histories is believed to be a direct 
indication of the iron production enhanced by an increasing amount of type Ia SNe (SNe~Ia) 
from old stellar populations \citep{carney96, Kirby19}. 
However, in the case of GCs, it should be noted that according to the 
statistical studies by \citet{ch12}, the mean age of the metal-rich Galactic and M31 GCs is 10.637 $\pm$ 0.468 Gyr, 
and the mean age of the metal-poor Galactic and M31 GCs is 10.217 $\pm$ 0.226 Gyr. The high-metallicity group of 
Galactic GCs contains on average fifteen times greater amount of iron and twelve times greater amount of magnesium 
than that of low metallicity \citep{ASh18}.  These differences in the  abundances cannot be explained by the 
predominant contribution of core-collapse SNe  (SNe~CC$=$SNe~type~II and SNe types Ib/c)\footnote{SNe~CC explode on time scales of 
$\sim 10^{6-7}$~yr and enrich the environment primarily with the O and Mg elements. The $\rm \alpha$-process elements, 
such as Si, S, Ca, and Ti, are produced during the explosions of 
one SN~Ia in comparable and even larger quantities than those produced in the explosion of one SN~CC. 
This follows from the SNe~Ia models by \citet{Tsujimoto95} and \citet{LeungNomoto18}. } 
to nucleosynthesis. 
These facts can be naturally explained thanks to the following recent theoretical discoveries of the last decade. 
These studies have shown that SNe~Ia are formed not only in old stellar populations but also in young stellar 
populations \citep{Bartunov94, Mannucci05, Li11, Rigault13, Kim18}. That is, not only SNe~CC but also SNe~Ia 
from short-lived progenitors explode in GCs at early evolutionary epochs. The nature of the short-lived progenitors 
of SNe~Ia has not yet been established. A young or so-called prompt progenitor population produces SNe~Ia 
on timescales of $\rm 100-330$~Myr (e.g., \citet{Aubourg08, MaozBadenes10} and references therein).  

 The analysis of the SNe~Ia spectra \citep{Nomoto13} suggests that, most likely, each such event is an explosion 
of a white dwarf that occurs as a result of its interaction with the matter of the companion star.
Since the density of stars in globular clusters is high, it can be expected that the interaction of 
stars in them will occur more often than in other stellar groups. Interacting binaries experience 
significant deviations from a single star evolution.
The evolution of multiple stars allows, at least in one of them, a carbon-oxygen core, capable of 
a subsequent explosion, with a mass of about $\rm 1.4 M_{\sun}$ required by the theory to form faster 
in comparison with that of a single star \citep{anguiano20}. 
On this basis, we believe that GCs are the most appropriate laboratories for studying the 
Prompt SN~Ia nucleosynthesis.
 
 Analysis of the fine features in the chemical pattern of the Galactic disc in the paper by \citet{AM13} 
 has shown that there is approximately three-times difference in the mass between the synthesized Fe per a SN~Ia  
 exploding in the star-formation regions and a SN~Ia  exploding in the old stellar population: 
 $\rm 0.23\pm 0.06 M_{\sun}$ and  $\rm 0.6\pm 0.2 M_{\sun}$, respectively. 
 A value of $\rm 0.6 M_{\sun}$ coincides with that obtained in the calculations of nuclear burning in 
 a white dwarf (WD) in the classical SN~Ia models: 
 W7 -- pure turbulent deflagration models and WDD -- delayed detonation or deflagration-detonation 
 transition models \citep{Nomoto84, Thielemann93}. These models both belong to the so-called  ``Single Degenerate'' type. 

Theoretical possibility of producing a reduced number of iron peak elements during a SN~Ia explosion was first 
found by \citet{LeungNomoto18} in the study of various burning regimes in white dwarfs. 
If carbon explodes in the centre of a WD with a low central density from $\rm 0.5 \cdot10^9 g/cm^3$ to $\rm 1.0 \cdot10^9 g/cm^3$, 
then, as calculations show, the iron mass synthesized in 
a two-dimensional pure turbulent deflagration model can be three times lower than that obtained with the W7 and WDD models. 
According to the authors, the possibility of the carbon ignition at such low densities is possible with a 
powerful influx of matter onto the WD surface.  At a density of $\rm 0.5\cdot10^9 g/cm^3$, an iron mass 
of $\rm 0.21  M_{\sun}$ is synthesized \citep{LeungNomoto18}. With the density increase, the mass of the 
synthesized iron increases. At a density of $\rm 10^9 g/cm^3$, an iron mass of $\rm 0.265 M_{\sun}$ 
is synthesized \citep{LeungNomoto18}. A value of $\rm 0.23\pm 0.06 M_{\sun}$ for the iron nucleosynthesis in 
a Prompt SN~Ia obtained in the paper by \citet{AM13} is very close to their arithmetic mean of the aforementioned values.

Thus, the study of nucleosynthesis in GCs allows one to choose the Prompt SN~Ia progenitor model or put 
independent constraints on the theory of the SN nucleothynthesis and can help one understand 
the mechanism leading to formation of SNe~Ia from short-lived progenitors.
This is the subject of the present paper.

  The idea of our method can be shortly formulated as follows. 
  We will use the method of determination of 
  the Mg and Fe mass synthesized by  SNe~CC and Prompt SNe~Ia explosions in the low-metallicity subgroup of GCs, 
  described  in detail by \citet{ASh18}. In this way, we can estimate the corresponding number of SNe~CC and SNe~Ia. 
  Then, using the determined number of SNe of different types one can explain the nucleosynthesis of other chemical 
  elements. Finally, we choose the Prompt SN~Ia progenitor model. Nucleosynthesis of several chemical elements in this
model should most closely match the abundances of these elements in GCs obtained from the analysis.

\section[]{Observed data analysis}
 
Looking ahead, it should be noted that to study the properties of supernovae using our method,
described in detail below in Sec.~3,  {\it accurate} estimates of the abundances of several chemical elements are needed.
Such data available today {\it only} for GCs of our Galaxy will be analysed in this section.
We will begin our consideration of the observed data with the metallicity distributions of CGCs in order to 
emphasize the similarities in the metallicity distributions of CGCs and GCs and to argue for the scenario 
of CGC evolution associated with the formation of GCs. 
The proposed scenario is described in Sec.~\ref{scenario}.
%

Supernovae that exploded in low-metallicity GCs enriched the surrounding gas with metals. 
In order to correctly estimate the number of supernovae formed, it is necessary to understand 
the properties of the gas that is parental for GCs.

The literature data on metallicities and chemical abundances considered in this paper will 
be statistically analysed and illustrated using histograms.
The bins of the histograms are equal to 0.25~dex which corresponds to the mean error of 
the metallicity determination with these data. 

We will consider the metallicity distributions of different types of CGCs at different redshfts: 
at the epoch of the formation of the main amount of GCs, that is at $\rm 2 <z <4$ and for earlier epochs 
($\rm z <1$), which in the literature are associated with the formation of disk structural components of 
galaxies (e.g. \citet{ZasSil10} and references therein). The star-formation rate in them was high. 
Hence, it follows that the rate of supernova explosions
was also high, and the enrichment of the interstellar gas with heavier elements was efficient. 
The question is pertinent here: could galactic fountains have influenced the properties of CGCs and 
how did this happen?
The studies by \citet{Afruni21} clearly show, that supernova-driven outflows from the central galaxies most 
 likely have a much smaller role in the dynamics and origin of the cool circumgalactic medium than generally believed. 
 The main conclusion drawn by \citet{Afruni21, Afruni19} was the following. Most of the cool circumgalactic
medium is originated by the accretion of gas from the intergalactic medium.
We will compare the metallicity distributions for the redshift intervals $\rm 2 < z < 4 $ and $\rm z \le 1$
to check the influence of the disc stage of the galactic evolution on the metal abundance in CGCs.

Table~1 gives the types of the considered gaseous clouds, for  high- and low-metallicicty subgroups: 
the average values of the metallicity, the root-mean square deviations, the number of clouds, and references. 

 The statistics of the element abundances given in Table~1 is performed in relation to the dividing line $\rm [X/H]= -1$,
regardless of whether the histograms plotted from the literature chemical abundances show a clear division into 
high- and low-metallicity subgroups. To perform statistical evaluation,
we only require that data exist separately for metallicities $\rm [X/H]< -1$ 
(low metallicity subgroup) and $\rm [X/H]> -1$ (high metallicity subgroup).
As it was noted in the Introduction, the dividing 
line $\rm [X/H]> -1$ is associated in our study with the position of the minimum on the GC metallicity distribution \citet{ASh18}.

\subsection{Analysis of metallicities of circumgalactic clouds}

\begin{figure}
\includegraphics[scale=1.2]{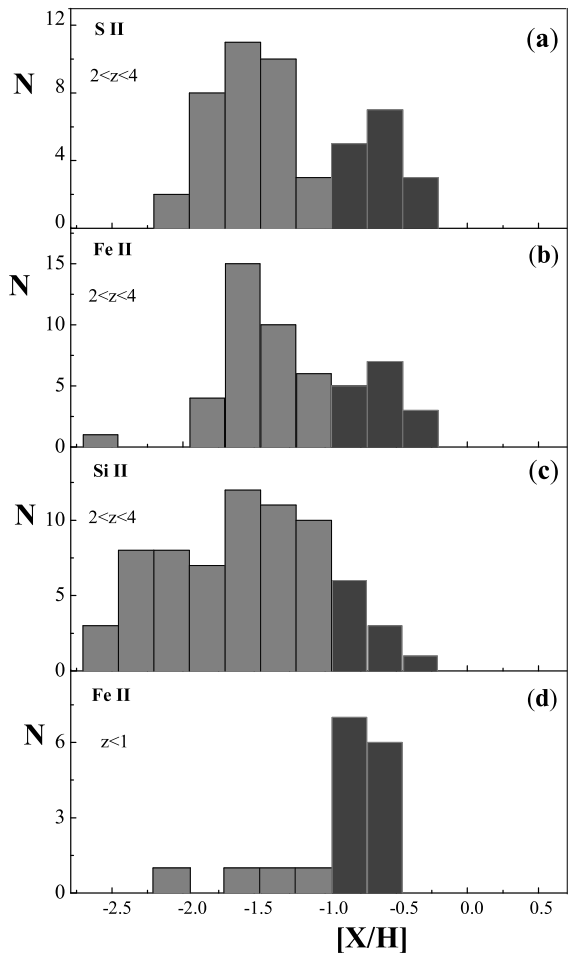}
\caption{Metallicity distribution of DLAs separately for each indicator:
S~II (panel a), Fe~II (panels b,d) ,Si~II (panel c) according to the data of \citet{Raf12}. 
The data shown for different redshift ranges: at the epoch of the formation of the major amount of GCs, that is at $\rm 2 <z <4$
(panels a--c) and at the disc stage of the galactic evolution $\rm z <1$ (panel d).
The high-metallicity cloud group is marked with dark-grey ($\rm [X/H] \ge -1$), the low-metallicity cloud group is marked with grey.}
\label{fig00}
\end{figure}

\begin{figure}
\includegraphics[scale=1.2]{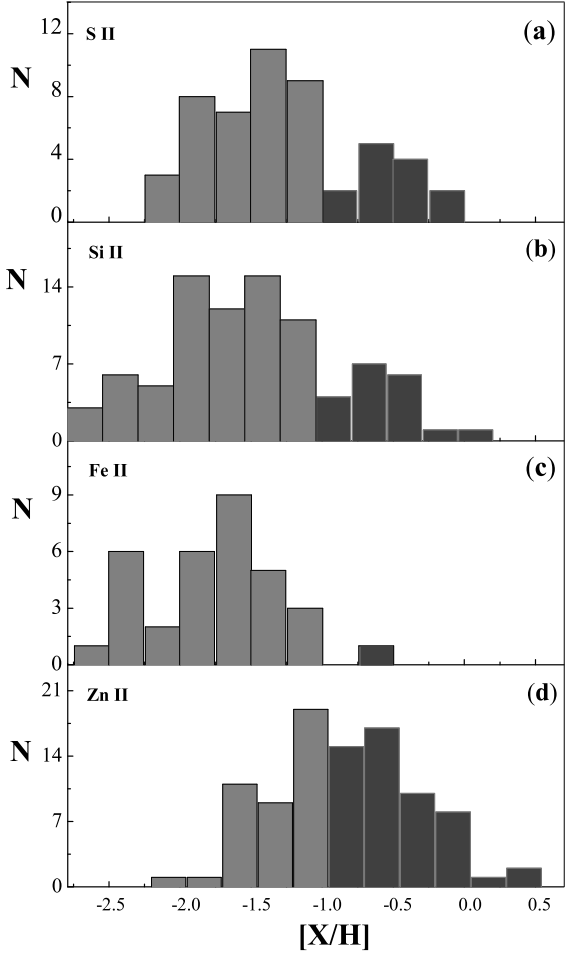}
\caption{Metallicity distribution of sub-DLAs separately for each indicator:
S~II (panel a), Si~II (panel b), Fe~II  (panel c), Zn~II (panel d) according to the data of \citet{Q16}. 
The data in all the panels is shown for the redshift range of  $\rm 1.8 <z <4$. 
The high-metallicity cloud group is marked with dark-grey ($\rm [X/H] \ge -1$), 
the low-metallicity cloud group is marked with grey.}
\label{fig2}
\end{figure}

\citet{ASh18} discussed the possibility to consider partial Lyman limit 
systems and Lyman limit systems as the residual parts of clouds, in which GCs were formed. 
The conclusion was drawn from the statistical analysis of the Mg abundances in GCs \citep{d16,carretta10,pritzl05} 
and in CGCs \citep{w16}, and on the spatial location of these objects.  
The analysis was carried out for the CGCs located at the redshifts $\rm z \le 1$. 
Let us consider the metallicity distribution of CGCs 
at the epoch of the formation of the major amount of GCs, that is at $\rm 2 <z <4 $ 
which corresponds to a time interval of about 9 to 12~Gyr ago. 

 Gaseous clouds within the virial radii of galaxies are usually classified according 
 to their neutral hydrogen column densities ($\rm  lgN_{HI}$) as follows: 
 damped Lyman limit systems (DLAs) with  $\rm lgN_{HI} \ge 20.3$, 
 sub-damped Lyman limit systems (sub-DLAs) with $ \rm 19\le lgN_{HI}<20.3$,   
 Lyman limit systems (LLSs) with $\rm 17.2 \le lgN_{HI}<19$, and partial    
 Lyman limit systems (pLLSs) with $\rm 16.1\le lgN_{HI}<17.2$ \citep{w18, Q16}.  
 The lower $\rm N_{HI}$ is, the higher the gas ionization rate is.
DLAs and sub-DLAs are mainly composed of neutral gas. pLLSs and LLSs are almost completely ionized \citep{Lehner16}.
    While studying the location of CGCs relative to galaxies, one should take into account that only most luminous 
    galaxies can be observed at redshifts  higher than $ \rm z \sim 2$. 
    In relation to such galaxies, 
   pLLSs and LLSs are always found within a virial radius of $ \le 300$~kpc with more than a half found 
   in the circumgalactic medium of massive galaxies \citep{Rudie12}. It cannot be ruled out that all pLLSs 
   and LLSs may belong to the circumgalactic medium of less massive galaxies at high z that cannot be detected  
   \citep{Lehner16}. DLAs at $ \rm z \sim 3 $ are observed in the outer regions of galaxies \citep{Raf11}.  
   At the redshifts $\rm  z <1 $,  pLLSs and LLSs are located presumably within the galactocentric radii    
   $\rm R_{gc}\sim 30 - 100$~kpc, but DLAs are distributed within $\rm R_{gc}<20$~kpc \citep{l13, Fum16}.

\subsubsection{DLAs and sub-DLAs}

We will start our analysis of the metallicity distribution with the densest cloud types: 
DLAs and sub-DLAs that consist mainly of neutral gas (see Fig.~\ref{fig00}). 
 Their metallicities are usually determined from the lines of singly ionized ions \citep{Raf12, Q16, w18}.  
 In \citet{Raf12}, Fe~II ions and the ions of the $\rm \alpha$-process elements (S~II, Si~II) 
were used to determine metallicity for $\rm 2 <z <4$. 
Sulphur is known to be a volatile element with negligible depletion. 
This element is considered the best option for determining the metallicity composition.
Therefore, if it is possible to determine the composition from this ion, then it is considered that 
$\rm [X/H]=[S/H]$, and other indicators are not used in \citet{Raf12}.
Si is the second choice element, it is used if the metallicity cannot be determined from the lines 
of S~II. It is assumed that $\rm [X/H]=[Si/H]$. For all the clouds, the metallicity was determined 
by \citet{Raf12} simultaneously on the basis of Fe~II ions, $\rm [X/H]=[Fe/H]+0.3$.
Note that the histograms obtained from different metallicity indicators are different.

The metallicity distribution, built from S~II, shows a division into high- and low-metallicity subgroups, 
with two well-defined maxima and cloud deficit near ($\rm [X/H]\approx-1$) (Fig.~\ref{fig00}, panel \textbf{a}). 
The high-metallicity subgroup has half the amount of clouds than the low-metallicity subgroup (see Table~1). 
The average metallicity of the low-metallicity clouds ($\rm [X/H] <-1$) is 
$\rm \langle [X/H] \rangle = -1.53 \pm 0.26$, and for the high-metallicity clouds 
($\rm [X/H]> -1$), it is $\rm \langle [X/H] \rangle = -0.61 \pm 0.19 $.
The metallicity of the same clouds, determined from the Fe~II ion abundance, shows a less unambiguous division 
into high- and low-metallicity subgroups (Fig.~\ref{fig00}, panel \textbf{b}). The high-metallicity peak is 
much weaker than that obtained in the previous case.
The average metallicity of the low-metallicity clouds ($\rm [X/H] <-1$) is $\rm \langle [X/H] \rangle = -1.47 \pm 0.22$, 
 and for the high-metallicity clouds ($\rm [X/H]> -1$), it is $\rm \langle [X/H] \rangle = -0.64 \pm 0.31$. 
90\% of other 69 clouds from the \citet{Raf12} sample in the same redshift range, the metallicities of which 
were determined based on Si~II ions (Fig.~\ref{fig00}, panel \textbf{c}) belong to low-metallicity subgroups 
with $\rm \langle [X/H] \rangle = -1.72 \pm 0.42$.

Figure~\ref{fig00} shows a noticeable difference in the distributions of $\rm [X/H]$ in S~II and Si~II ions 
with respect to the dividing line $\rm [X/H]= -1$.  
The minimum between the high- and low-metallicity components is distinct only in the distribution based on S~II ions,
which most reliably reflects the metallicity \citep{Raf12}.

\begin{figure}
\includegraphics[scale=1.2]{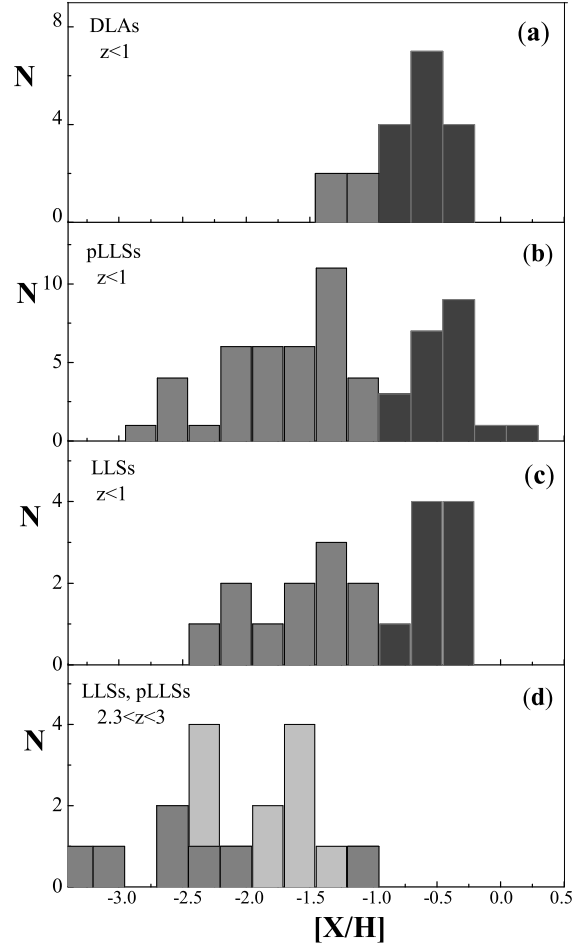}
\caption{Metallicity distribution of DLAs (panel a), pLLSs (panel b) and LLSs (panel c) for different redshift ranges  
according to the data from \citet{w18} (panels a--c) and LLSs and pLLSs according to \citet{Lehner16} (panel
d). The high-metallicity cloud group is marked with dark-grey ($\rm [X/H] \ge -1$), the low-metallicity cloud group is marked with grey.
Panel (d) shows pLLSs in light-grey colour.}
\label{fig3}
\end{figure}

In \citet{Raf12}, $\rm [X/H]$  for $z <1$ was determined, almost completely, from Fe~II ions. 
As can be seen from (Fig.~\ref{fig00}, panel \textbf{d}), this study presents clouds mainly from the high-metallicity subgroup. 
The average metallicity of the high-metallicity clouds ($\rm [X/H]> -1$) is $\rm \langle [X/H] \rangle = -0.74 \pm 0.13$ (see Table~1). 
As will be seen from further discussion (see also Fig.~\ref{fig3}), for pLLSs and LLSs at $z <1$, on the contrary, there are more 
low-metallicity clouds than high-metallicity ones. Therefore, the profile observed by \citet{Raf12} for DLAs metallicity requires 
further careful thought which is out of the scope of this study.

Next, let us consider the metallicity measurements of sub-DLAs located at $\rm 1.8 <z <4.0$ (see, please, Table~1, Fig.~\ref{fig2}).
These data were compiled from the literature by \citet{Q16}. 
The metallicity was determined using ions of different metals and also represented as a histogram with the averaging bin of 0.25~dex. 
We have extended the redshift range to $\rm z =1.8$, since half of the high-metallicity clouds, whose metallicity was measured based 
on S~II ions, were in the range of $\rm 1.8 <z <2.0$. Let us discuss the resulting distributions.

The metallicity distribution of 51 clouds obtained from the analysis of the S~II ion lines has the minimum near 
$\rm [X/H]= -1 $ and is divided into two components: the high-metallicity component (about 30\% of clouds) with 
$\rm \langle [X/H] \rangle = -0.54 \pm 0.26 $ and the low-metallicity component with $\rm \langle [X/H] \rangle = -1.53 \pm 0.33$ 
(Fig.~\ref{fig2}, panel \textbf{a}).

The metallicity distribution of the other 86 clouds, obtained from the analysis of the Si~II ion lines, is represented mainly 
by the low-metallicity component.
The high-metallicity component (about 20\% of clouds) is characterized by $\rm \langle [X/H] \rangle = -0.55 \pm 0.27$  
and the low-metallicity component -- $\rm \langle [X/H] \rangle = -1.73 \pm 0.42$ (Fig.~\ref{fig2}, panel \textbf{b}).

The metallicity distribution of 33 clouds, obtained from the analysis of the Fe~II ion lines, is represented by 
a low-metallicity component (97\% of clouds) with $\rm \langle [X/H] \rangle = -1.79 \pm 0.40$ (Fig.~\ref{fig2}, panel \textbf{c}).

In contrast to the result based on Si~II and Fe~II ions, 94 clouds, the metallicity of which was determined based on 
the analysis of the abundance of Zn~II ions, appeared to be mainly high-metallicity (about 63\% of clouds) with 
$\rm \langle [X/H] \rangle = -0.54 \pm 0.33$. The low-metallicity component is characterized by $\rm \langle [X/H] \rangle = -1.35 \pm 0.24$ 
(Fig.~\ref{fig2}, panel \textbf{d}). 
As well as in the case with the \citep{Raf12} data, only the distribution based on S~II ions demonstrates the minimum near
$\rm [X/H]=-1$.

Figures~\ref{fig00} and \ref{fig2} demonstrate that finding the dependence of the average metallicity of CGCs on redshift 
can give a deliberately false result if the determination of metallicity for different ranges of redshifts is based on different ions.
For example, Fig.~6 from \citep{Raf12} shows that the metallicity in the range of $\rm 2 <z <4$ was determined mainly from Si~II ions, 
and for $\rm z <1$, in most cases, from Zn~II ions. Figures ~\ref{fig00} and Fig.~\ref{fig2} demonstrate that the distribution 
determined from Si~II ions shows mainly a low-metallicity component of clouds, while that determined from Zn -- a high-metallicity component.

The metallicities of DLAs at $\rm z <1$ were determined by \citet{w18}(see,please, Fig.~\ref{fig3}, panels \textbf{a}). 
The metallicity indicators chosen by \citet{w18} are as follows: Zn~II for nineteen CGCs, Fe~II for nine CGCs, S~II for five CGCs, 
and Si~II for two CGCs. The authors used the following abundance ratios: 
$\rm [\alpha/Zn]\approx 0.0 \pm 0.2$, that is $\rm [X/H]=\znh$, $\rm [X/H]=[\alpha/H]$, and $\rm [X/H]=\feh +0.5$. 
 Only a component with $\rm [X/H]> -1$ can be seen, 
that is, 29 of 35 clouds (see Fig.~\ref{fig3}). 
The value $\rm \langle [X/H] \rangle = -0.57 \pm 0.26 $ coincides within the measurement errors 
with the mean value of the high-metallicity component of the distribution from \citet{Raf12} at $\rm 2 <z <4$, 
and from \citet{Q16} at $\rm 1.8 <z <4$. 
It likely means (see Table~1, sixth column) that the changes of the metallicities of clouds during the evolution from $\rm z=4$ 
up to present time are not detected for the high-metallicity subgroup.

\subsubsection{LLSs and pLLSs}
 LLSs and pLLSs are almost completely ionized.
 The gas in LLSs and pLLSs is often multiphase with the absorption seen in different ionization stages.

For pLLSs and LLSs at $\rm z <1$, the resulting sample of CGCs in \citet{w18} is the following: 82 pLLSs, 29 LLSs.  
 The metallicity of each cloud was determined by analysing the abundance of several ions: Mg~II, O~I, O~II, S~III, Fe~II, Si~III and Si~IV. 
Using the observed data, we excluded the objects with uncertain abundances from the sample according to the data from Table~6 in \cite{w18}.
The excluded clouds have the following metallicities: $\rm [X/H]<-3$~dex and $\rm [X/H]>1$~dex. 
These are the objects having only lower or upper limit abundance estimates. The resulting sample of pLLSs and  LLSs  consists 
of 60 and 20 members 
respectively (see Fig.~\ref{fig3}, panels \textbf{b} and \textbf{c}). 
For the average metallicity of the low-metallicity and high-metallicity clouds obtained by \citet{w18} see, 
please, third and sixth columns in Table~1.

\citet{Lehner16} measured metallicities for sixteen pLLSs and seven LLSs at $\rm 2.3 <z <3.0$. 
The metallicity was determined mainly by ions at a high ionization level, Si~III and Si~IV. 
The data obtained by \citet{Lehner16} contain a very metal-poor component with $\rm [X/H] \le -2.3$ for LLSs 
(see Fig.~\ref{fig3}, panel \textbf{d}). This feature only slightly manifests itself in the histogram for pLLSs at $\rm z <1$ \citep{w18}. 
We do not consider it in this paper, 
because the GC samples used in this paper (see Sec.~2.2) do not show any analogue of this feature 
in their metallicity distribution.
\citet{cooke17} suggest that most metal-poor CGCs are progenitors of the lowest-mass galaxies. 
The study of this hypothesis is beyond the scope of this study. 
We only note that several GCs in this metallicity range are also known in our Galaxy and other galaxies \citep{harris,larsen}.

The average metallicity of pLLSs is $\rm \langle[X/H] \rangle = -1.69 \pm 0.20 $ \citep{Lehner16}.
The high metallicity component is not pronounced.

It is instructive to compare the $\rm [X/H]$ distribution for DLAs at high redshifts measured by 
\citet{Raf12} and \citet{Q16} and the corresponding distribution for pLLSs at $\rm z <1$ from \citet{w18}. 
The comparison shows that they have similar mean $\rm [X/H]$ values for the high- and 
low-metallicity components (see Table~1, third and sixth columns).

\begin{table*}
 \centering
\caption{ Metallicities of circumgalactic clouds used in the statistical analysis}
\begin{tabular}{l|c|ccc|cccc}
\hline \hline
  &  &  \multicolumn{3}{c}{Low-metallicity subgroup } & \multicolumn{3}{c}{High-metallicity subgroup} &  \\ %
  &  &  \multicolumn{3}{c}{($\rm [X/H] <-1$)}           & \multicolumn{3}{c}{($\rm [X/H] \ge-1$)}                 &    \\ \hline

objects &  redshifts  & $\rm <[X/H]> $ & $\rm \sigma([X/H])$ & number & $\rm <[X/H]>$ & $\rm \sigma([X/H])$ & number & reference\\ 
   & &           &      &  of clouds  &         &        &  of clouds    &     \\ \hline

DLAs, S~II   & $\rm 2<z<4$&-1.53 & 0.26 & 34 & -0.61 & 0.19 & 15 & \citet{Raf12} \\
DLAs, Fe~II   & $\rm 2<z<4$&-1.47 & 0.22 & 35 & -0.64 & 0.31 & 13 & \citet{Raf12} \\
DLAs, Si~II   & $\rm 2<z<4$&-1.72 & 0.42 & 57 & -0.81 & 0.16 & 12 & \citet{Raf12} \\
DLAs, Fe~II   & $\rm z<1$&-0.74 & 0.13 & 13 & -- & -- & -- & \citet{Raf12} \\
sub-DLAs, S~II & $\rm 1.8<z<4$ &-1.53 & 0.33 & 37 & -0.54 & 0.26 & 14 & \citet{Q16} \\
sub-DLAs, Si~II & $\rm 1.8<z<4$ &-1.73 & 0.42 & 68 & -0.55 & 0.27 & 18 & \citet{Q16} \\
sub-DLAs, Zn~II & $\rm 1.8<z<4$ &-1.35 & 0.24 & 35 & -0.54 & 0.33 & 59 & \citet{Q16} \\
sub-DLAs, Fe~II & $\rm 1.8<z<4$ &-1.79 & 0.40 & 32 & -- & -- & -- & \citet{Q16} \\
DLAs & $\rm z<1$ & -- & -- & -- & -0.57 & 0.26 & 29 &\citet{w18} \\
pLLSs & $\rm z<1$ & -1.75 & 0.46 & 37 & -0.50 & 0.27 & 23 & \citet{w18}\\
LLSs & $\rm z<1$ & -1.60 & 0.42 & 11 & -0.45 & 0.19 & 9 & \citet{w18} \\
pLLSs & $\rm 2.3<z<3$ & -1.69 & 0.20 & 8 & -- & -- & -- & \citet{ Lehner16} \\
\hline
\end{tabular}
\label{tab:wotta18}
\end{table*}

We can summarize the properties of the clouds as follows. 
 All types of CGCs (DLAs, sub-DLAs, LLSs and pLLSs) are observed at the redshifts $\rm z < 4 $.
The average metallicity within each subgroup (high- and low-metallicity) coincides for pLLSs, LLSs, and DLAs within the 
estimated errors for the redshift intervals considered in Table~\ref{tab:wotta18}. This implies that the evolution of the metallicity 
in the high- and low-metallicity components of CGCs, separately, is not evident starting from $\rm z \sim 4 $, 
and different types of clouds for a given metallicity show similar mean [X/H] values within 
the corresponding errors of their determination.


\subsection{ Comparison of the metallicities of GCs and CGCs}
In this section, we use the results of the statistical study by \citet{ASh18}, who have analysed the $\rm \mgh$ 
and $\rm \feh$ distributions of Galactic GCs.

\begin{figure}
\includegraphics[scale=0.8]{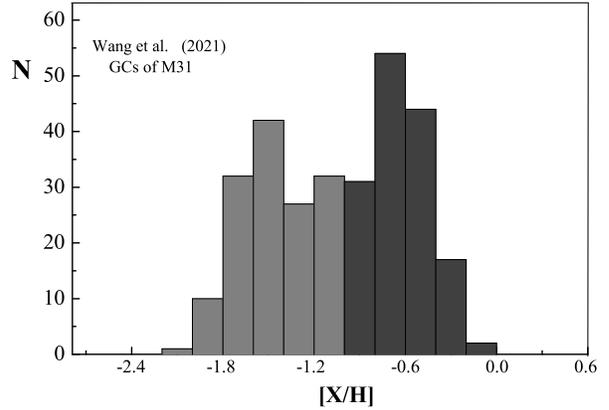}
\caption{Distributions of abundances of $\rm [X/H]=[\alpha/H]$ in GCs built using the data from \citet{d16} 
for the objects of three Galactic subsystems: the disc, the inner halo, and the outer halo.}
\label{fig4}
\end{figure}

As in the case of the clouds (Sec.~2.1), the $\rm [X/H]$ 
values for GCs were determined in the literature by analysing different elemental abundances. 
Figure~\ref{fig4} shows the distributions of $\rm \alpha$-element abundances in GCs built using the data of the integrated-light
spectroscopy from \citet{d16} for the objects in three subsystems of the Galaxy: the disc, the inner, and outer halos. 
\citet{ASh18} plotted the $\rm \feh$ and $\rm \mgh$ distributions using the data from the MILES library \citep{Sanchez06}.  
Since the [X/H] metallicity is determined using $ \rm \alpha $-elements Mg, Ca, and Ti, 
we need to know the $ [\rm \alpha/Fe] $  values for GCs.
For this, we use the data from \citet{d16} obtained from the Coelho library \citep{Coelho05}. 
This is the largest homogeneous data sample of $\rm \alpha$-element abundances to date. 
We avoid to compile abundances from various literature sources, because they may lead to wrong conclusions 
due to possible systematic deviations in the data of different authors (see, e.g., \citet{Schiavon12}). 
The distributions of $\rm [\alpha /H]$  and $\rm \mgh$ \citep{d16} look very similar (Fig.~\ref{fig4}).

 Let us consider the value  $\rm [X/H]=-1$ as the boundary between the low- and high-metallicity subgroups 
 of GCs as in the case of CGCs. Hereafter, we will provide statistical analysis based on this division. 
 Table~\ref{tab:dias16} shows the number of GCs in each subgroup, the average  $\rm [X/H]$  values and 
 the root-mean-square deviations according to \citet{d16} obtained from the Coelho library. 

\begin{table*}
 \centering
\caption{Average  $\rm [X/H]$  values,  root-mean-square deviations, and
the number of GCs according to \citet{d16} for our Galaxy and \citet{Wang21}
for the M31 galaxy}
\begin{tabular}{|l|ccc|ccc|}
\hline \hline
 &  \multicolumn{3}{c}{Low-metallicity subgroup $\rm [X/H] <-1$ } & \multicolumn{3}{c}{High-metallicity subgroup $\rm [X/H] \ge-1$} \\  \hline

   Galactic subsystem     & $\rm \langle [X/H] \rangle$ & $\rm \sigma([X/H])$ & number of GCs& $\rm \langle[X/H]\rangle $& $\rm \sigma([X/H])$ & number of GCs \\ \hline
disc & -1.65 & 0.30 & 6 & -0.59 & 0.17 & 12 \\
inner halo & -1.74 & 0.43 & 8 & -0.69 & 0.24 & 5 \\
outer halo & -1.62 & 0.28 & 8 & -- & -- & -- \\    \hline

   M31 galaxy  & -1.44 & 0.25 & 145 & -0.63 & 0.19 & 148 \\

\hline \hline
\end{tabular}
\label{tab:dias16}
\end{table*}

\begin{table*}
 \centering
\caption{Abundances of chemical elements in GCs determined using their integrated-light spectra by \citet{colucci17}}
\begin{tabular}{l|ccc|ccc}
\hline \hline
chemical        &  \multicolumn{3}{c}{\underline{Low-metallicity subgroup ($\rm [X/H] <-1$)} } & \multicolumn{3}{c}{\underline{High-metallicity subgroup ($\rm [X/H] \ge-1$)}} \\  
 element   &  $\rm \langle [X/H] \rangle $ &  $\rm \sigma([X/H])$ & $\rm N_{GC}$ &  $\rm \langle [X/H] \rangle $ &  $\rm \sigma([X/H])$ & $\rm N_{GC}$  \\  \hline \hline
\smallskip
        &\multicolumn{3}{c}{$\rm \mgh<-1.0$ } &         \multicolumn{3}{c}{$\rm \mgh \ge -1.0$ } \\ 
$\rm^{24}$Mg     & -1.57 & 0.40 & 4 & -0.38 & 0.26 & 7 \\ \hline
\smallskip
  &\multicolumn{3}{c}{$\rm \sih<-0.7$ } &         \multicolumn{3}{c}{$\rm \sih \ge -0.7$ } \\  
$\rm^{28}$Si     &-1.14 & 0.25 & 3 & -0.18 & 0.19 & 7 \\ \hline
\smallskip
  &\multicolumn{3}{c}{$\rm \cah<-0.8$ } &         \multicolumn{3}{c}{$\rm \cah \ge -0.8$ } \\ 
$\rm^{40}$Ca  & -1.34 & 0.35 & 4 & -0.32 & 0.21 & 7 \\  \hline
\smallskip
  &\multicolumn{3}{c}{$\rm \crh<-1.0$ } &         \multicolumn{3}{c}{$\rm \crh \ge -1.0$ } \\
$\rm^{52}$Cr & -1.68 & 0.41 & 4 & -0.57 & 0.25 & 7 \\ \hline
\smallskip 
 &\multicolumn{3}{c}{$\rm \feh<-1.0$ } &         \multicolumn{3}{c}{$\rm \feh \ge -1.0$ } \\ 
$\rm^{56}Fe$  & -1.63 & 0.35 & 4 & -0.48 & 0.21 & 7 \\ 
\hline \hline
\end{tabular}
\label{tab:colucci17}
\end{table*}

The coincidence within the errors of the $\rm \langle \mgh \rangle$ mean values and the root-mean-square deviations 
for low- and high-metallicity subgroups of CGCs and GCs has allowed \citet{ASh18} to hypothesize that these clouds can 
be the residual parts of the clouds, in which GCs have been formed. 
 The considered here $\rm [X/H]$ data by \citet{w18, Raf12, Q16, Lehner16} argue in favour of the conclusions by \citet{ASh18}. 
The average metallicity of 
the low-metallicity Galactic GC subgroup $\rm \langle [X/H] \rangle =-1.67 \pm 0.35$, whereas the average metallicity of 
the high-metallicity Galactic GC subgroup $\rm \langle [X/H] \rangle =-0.62 \pm 0.25$. 
The average metallicity of the subgroup 
of low-metallicity CGCs $\rm \langle [X/H] \rangle =-1.61 \pm 0.39$, whereas the average metallicity of the 
high-metallicity  CGC subgroup $\rm \langle [X/H] \rangle =-0.52 \pm 0.28$.

\begin{figure}
\includegraphics[scale=0.8]{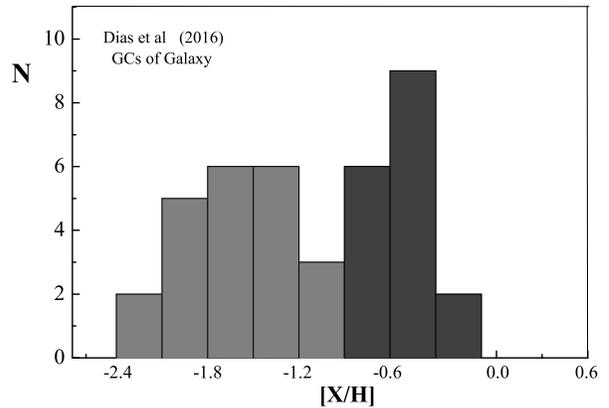}
\caption{Distributions of the $\rm [X/H]$ abundances in GCs of the M31 galaxy built using the data from \citet{Wang21} 
for the old ($\sim$10 Gyr) objects of three Galactic subsystems: the disc, the inner halo, and the outer halo.}
\label{fig5}
\end{figure}

In this regard, it is important to quote the studies of metallicity distributions for GCs in other galaxies. 
By combining the information of the LAMOST spectra and the multi-band photometry, \citet{Wang21} 
determined the ages of GCs and derived parameters of 53 young and 293 old clusters in the Andromeda galaxy. 
Most old clusters have the ages approximately equal to 10~Gyr. 
The metallicity distribution of $\rm [X/H]$ for old GCs in the M31 galaxy is shown in Fig.~\ref{fig5}. 
It is similar to the metallicity distribution for GCs of our Galaxy. 
 The high- and low-metallicity peaks of the distribution are less pronounced, but there is a natural explanation for this fact.
The uncertainties of the metallicity estimates and the metallicities of GCs originated in the processes of active
galactic interactions \citep{Wang21} can wash away sharp features. The average metallicity of the subgroup 
of the low-metallicity GCs $\rm \langle [X/H] \rangle =-1.44 \pm 0.25$,
whereas the average metallicity of the high-metallicity subgroup of GCs $\rm \langle [X/H] \rangle =-0.63 \pm 0.19$. 
The number of GCs in each group is given in Table~2. 

 Large galaxies in the centres of galactic clusters have experienced
more complex star-formation histories with violent star-forming events than faint isolated galaxies 
(e.g., Kruijssen et al. 2019, Villaume et al. 2020, Longobardi et al. 2015). Interestingly, even in such massive galaxies,
statistical analysis is able to reveal $\sim$10-Gyr old metal-rich and metal-poor GC populations  
with the corresponding metallicity peaks similar to the ones considered in this section \citep{das15}.
For example, \citet{peng06} and \citet{Harris09} have studied the extragalactic GC systems in giant elliptical galaxies and 
argued that the metallicity distributions of GCs in them are bimodal.

Abundances of chemical elements in extragalactic GCs obtained using their integrated-light spectra 
are determined together with their ages using stellar population models.
Even in the galaxies and galaxy subsystems with prevailing old stellar populations, there may be a significant population of
intermediate age GCs ($\rm 1 \le T \le 7$~Gyr). For example, about 15 \% of GCs are $\le 3$ Gyr old in M31  \citep{Wang21}. 
These abundances can also
be biased (see, e.g., \citet{Schiavon12}) due to the differences in the methods of the observed data reduction 
and analysis used by different authors and due to the low signal-to-noise ratios in the spectra of distant objects.
We, therefore, used only the data for Galactic GCs from \citet{d16} for the analysis in this section. 
\citet{d16} demonstrated the agreement of their estimates with the literature data.  

\subsection{Possible scenario of CGC evolution associated with the formation of GCs} 
\label{scenario}

 Before we consider the properties of supernovae responsible for the enrichment of GCs and their parent CGCs with chemical 
elements, let us discuss formation of GCs.
Since the average ages of the high- and low-metallicity subgroups of GCs discussed above are similar \citep{ch12, vandb13}, 
 the metal enrichment of the low-metallicity clouds has happened fast as a result of explosions of massive stars or rapidly 
 evolving binary stellar systems. As it was shown in \citet{ASh18}, the fraction of mass of the enriched part of the cloud 
 is from 20\% to 50\%  of the initial cloud mass depending on the fraction of the cloud gas transformed into stars. 
A high-metallicity group of GCs is formed from this enriched gas. The analysis of the observed data for CGCs and GCs, 
performed in this paper, identified several phenomena related to our study, the further study of which can add important 
refinements to the theory of galaxy formation and evolution.

There are very metal-poor clouds shown in Fig.~\ref{fig3} with $\rm \langle [X/H] \rangle < -2.3$.
There are relatively few such clouds found. Some authors considered them as the gaseous clouds left over 
from the explosion of the first generation of stars \citep{Welsh19, Kulkarni13}. 
Can we associate these CGCs with GCs using the criterion for the closeness of their metallicities? 
Yes, we can. Although, we should admit, such GCs are very rare. \citet{beasley19} compiled spectroscopic metallicity data for the GC systems 
of 28 nearby galaxies and found 1--2 GCs with $\rm \langle [X/H] \rangle < -2.3$ only in 5 galaxies. 
In the paper by \citet{Kruijs19}, a possible explanation for this phenomenon is proposed, 
which is that the galaxies with metallicities of $\rm [Fe/H]\le-2.5$ have too small masses to form GCs with initial 
masses greater than $\rm 10^{5} M_{\sun}$ and needed to survive for the Hubble time. 
The processes of enrichment with chemical elements of most metal-poor clouds require separate careful consideration.

The next special feature worth noting is as follows. Note that both on histograms for $\rm 2 <z <4$ and on histograms 
for $\rm z <1$ (Fig.~\ref{fig00} and Fig.~\ref{fig2}), there are practically no CGCs with $\rm [X/H] \ge -0.3$. 
Figures~\ref{fig4} and ~\ref{fig5} show that a similar situation is observed for GCs. GCs formed from high-metallicity clouds, 
in turn, should have contained supernovae that enriched the surrounding gas. 
Therefore, the absence of very high-metallicity GCs and CGCs requires an explanation. 
With care, one can tentatively assume that cloud fragments enriched to metallicity $\rm [X/H] > -0.3$ pass into another 
phase due to rapid cooling on metals; and the conditions are created in them for the clusters with $\rm M < 10^{5} M_{\sun}$ 
to form, which do not survive for the Hubble time according to \citet{Kruijs19}.

A natural question is why we do not observe the formation of GCs in the near-galactic clouds in the present epoch.
\citet{mandelker18} presented a new model of the GC formation and showed that extremely turbulent conditions are 
required for cold filamentary accretion to lead to the formation of star-forming clumps. 
According to calculations, the densities required for the formation of GCs are achieved with 
the collision of counter-rotating streams of very massive clouds. Such conditions were likely to occur at redshifts 
of $\rm  z> 4.5 $ \citep{mandelker18}. Therefore, it must be assumed that GCs, which formed from CGCs, are of 
about 10 billion years old or older. GCs can also form in dwarf galaxies. In this case, it is natural to assume 
that the metallicity of these GCs will correspond to the metallicity of the CGCs.
Therefore, it is of interest to study the chemical composition of GCs in isolated low-mass dwarf galaxies which 
probably did not undergo merger processes with other galaxies \citep{sharina17}.

At present, the collision of clouds resulting in the formation of star clusters, the masses of which 
are comparable to the initial mass of GCs,  $\rm 10^6-10^7 M_{\sun}$, can be found only in the processes of major merging. 
Antennae Galaxies NGC 4038/NGC 4039 \citep{tsuge21} are a good example. The collision relative velocity of the clouds 
in these galaxies, according to estimates, is about $\rm 100 km/s$. It is natural to expect that the metallicity 
of GCs formed in such processes will be equal to the metallicity of gas, that has undergone several stages 
of transformation by stars, and will differ from the metallicity of CGCs that existed at the stage of the initial formation of galaxies.


\section[]{Selection of the proper SN~Ia model}


\subsection{Details of the method}

\begin{figure*}
\includegraphics[scale=0.63]{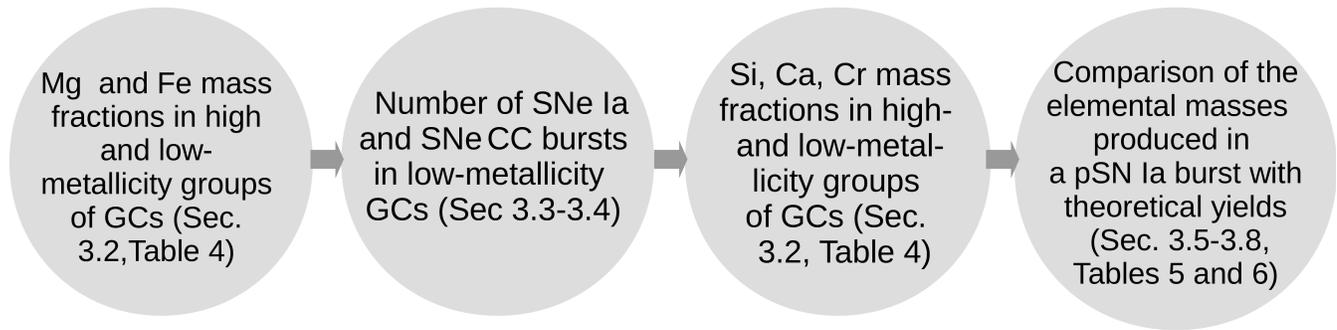}
\caption{Diagram explaining the method for selection of the proper Prompt SNe~Ia model.}
\label{fig0}
\end{figure*}
To select the proper Prompt SN~Ia model from the analysis of nucleosynthesis presented in Sec.~2, 
we need  accurate data on the abundances of chemical elements, the main production channels of which are 
two subtypes of SNe: SNe~CC and SNe~Ia. Note that the contribution of these elements from other stages of 
the stellar evolution should be negligible.
 The idea of our method can be shortly formulated as follows (please, see also Fig.~\ref{fig0}). 
 We will use the method of determination of the Mg and Fe mass, which has been synthesized by  SNe~CC and Prompt SNe~Ia 
explosions in the low-metal subgroup of GCs, which is described  in detail by \citet{ASh18}, section 4. 
In this way, we can estimate the corresponding number of SNe~CC and SNe~Ia. Then, using the determined number of SNe 
of different types, one can explain the nucleosynthesis  of other chemical elements.
 \citet{ASh18} show and we demonstrate in this paper that our calculations are independent of the fraction of the enriched cloud mass.
 We consider that the average mass of an element ejected into the interstellar medium during a single burst  of a SN~CC as known.
 This assumption is justified, because theoretical models of the SN~CC nucleosynthesis are confirmed by 
 direct observations of massive stellar progenitors of SNe~CC and by the analysis of the spectra of their expanding shells of 
different masses (\citet{smartt09} and references therein). The average mass of the element synthesized during the 
 SN~CC explosion is determined using formula (1) by \citet{Tsujimoto95} with the calculations by \citet{Nomoto06}. 
 The results of \citet{Tsujimoto95} and \citet{Nomoto06} almost coincide for the chemical elements under consideration.
It is necessary to clarify the maximum mass of stars involved in the enrichment process. Studies of the SN progenitors
 and their stellar remnants in nearby galaxies argue that the maximum mass of the progenitor of a SN~CC 
 is $\rm \sim 20 M_{\sun}$. Stars with masses higher than this limit experience implosion and form a black hole without any 
 release of the enriched substance (\citealt{AuchetteLopez18} and references therein). At the same time,  the authors conclude
 that there may be ``islands of explodeability''\footnote{The words cited from the paper by \citet{AuchetteLopez18} mean that 
 the masses of exploding stars ($\rm \ge 20 M_{\sun}$) can be described by a discrete mass function.
 This assumption is introduced for setting the theoretical limit on the maximum mass of a star that can participate 
 in the enrichment processes \citep{Heger03, Sukhbold16} in agreement with the observation estimates obtained from 
 studies of supernova remnants and their environments in star-forming regions \citep{smartt09, Sarbadhicary17}. 
 According to \citet{AuchetteLopez18}, stars
with masses up to $\rm 40 M_{\sun}$ can explode in these environments.}
  for stars heavier than the limiting mass. They used the processes in the Small  and Large Magellanic Clouds as the example. 
  In this case, even high-mass progenitors up to $\sim 40 M_{\sun}$ may explode 
 in the case of some specially selected physical parameters of stars and their multiplicity \citep{AuchetteLopez18}.  
 Since the maximum mass of stars depends on the mass of the parent molecular cloud \citep{Larson82} 
 and metallicity\footnote{It is believed that the masses of the first stars were even greater  than $\rm 40 M_{\sun}$.}, 
 it would be naturally  expected that stars of such great masses form in GCs.  When calculating the average masses of 
 newly synthesized chemical 
 elements, we will take into account the contribution from the progenitors of SNe~CC with masses from $ 8 $ to $\rm 40 M_{\sun}$. 
 The average mass of the synthesized oxygen significantly depends on the maximum mass of a star, 
 but it is not used in this study. The average masses of considered Mg, Si, Ca, Cr, and Fe are negligibly 
 dependent on whether we use $\rm 20 M_{\sun}$ or $\rm 40 M_{\sun}$  as the maximum mass of SNe~CC.

Let us discuss a series of SN~Ia models, with which we will compare the results of our calculations.
 As mentioned in the Introduction, they differ, 
first, in the explosion mechanism: pure turbulent deflagration models and models with deflagration-detonation 
transition. These two explosion mechanisms comprise the basis of all single degenerate models of SNe~Ia\footnote{
Another SN~Ia explosion channel is possible -- double degenerate -- the merger of two white dwarfs with masses 
close to that of the Sun. The double-degenerate scenario is possible in old stellar populations with ages larger 
than $ \sim 2.4~Gyr$. Therefore, we do not take it into account in this paper.}.

Second, the models differ in the mass of a degenerate carbon-oxygen WD. The W7 and WDD models are calculated for a dwarf
mass equal to the Chandrasekhar limit of $\rm  1.38 M_{\sun} $ \citep{Nomoto84}. 
In \citet{LeungNomoto18}, the models for the WD masses from $\rm  1.30 M_{\sun} $ to $ \rm 1.38 M_{\sun} $  are also considered.

Let us note that the average mass of a chemical element released during a
Prompt SN~Ia explosion is a free parameter of our theory for all 
elements except for iron, because there are independent studies consistent with  theoretical calculations of the explosive
nucleosynthesis of SNe~Ia  \citep{LeungNomoto18} only for iron \citep{AM13,Childress15}.

\subsection{Analysis of abundances of chemical elements in GCs}
\label{sec:analysis}

To find observation limitations on the nucleosynthesis, we use the results of studies of GCs, in which the abundances
of several chemical elements have been determined. Among them, those have been selected that are produced during SN bursts, 
while their nucleosynthesis does not depend on the metallicities of SNe. The following chemical elements meet our aims: 
Mg, Si, Ca, Cr, and Fe. Among the listed chemical elements, Mg shows the manifestation of chemical evolution in GCs 
(\cite{gratton19} and references therein).  The currently observed low-mass stars cannot reach 
the temperature threshold required for activating the Mg-Al conversion, thus, this nuclear burning must have occurred 
in more massive stars, already evolved and dead
\citep{gratton19}. Magnesium and oxygen (oxygen also experiences depletion) are indicators of the amount of SNe~CC in 
chemical evolution models, because about 98\%  of Mg  and $\rm O$ are produced by SNe~CC \citep{McWilliam}. 
At the same time, 
according to the method described in \citet{ASh18},
the depletion of magnesium in a certain fraction of stars does not influence the results of determination of
 the number of SNe~CC. The magnesium abundances in the low- and high-metallicity groups decrease by the same value.
 
  \citet{colucci17} measured the abundances of Mg, Si, Ca, Cr, and Fe in eleven GCs using one method and 
 the integrated-light high-resolution spectra of GCs. Seven of these GCs are in the metal-rich group and four GCs are 
  in the metal-poor group. According to the classification of \citet{carretta10}, nine of these GCs belong to the 
  disc or bulge and two GCs belong to the inner halo.

   In the \citet{pritzl05} sample, thirty-five low-metallicity 
  GCs belong mainly to the outer halo and six  metal-rich GCs belong to the disc or bulge. However, this sample 
  is not suitable for our method (see Subsec. 3.8 for details).   Apparently,  the  abundances of the brightest  stars in GCs,  
  given by \citet{pritzl05} cannot be representative for all their stars.
\begin{table*}
 \centering
\caption{Calculation of the mass fractions $\rm Z_{El}$ of Mg, Fe, Si, Cr, and Ca in  low-metallicity and high-metallicity subgroups of GCs.}
\begin{tabular}{llclclc}
\hline \hline
Element ($El$) & \multicolumn{2}{c}{\underline{values for Sun}}  & \multicolumn{2}{c}{\underline{Low-metallicity subgroup}} & \multicolumn{2}{c}{\underline{High-metallicity subgroup}} \\  
 (object)     &  $\rm lgZ_{El_{\sun}}$ & $\rm A_El$  & $\rm \langle [El/H] \rangle $  &  $\rm lg Z_{El}$ & $ \rm \langle [El/H] \rangle $ &  $\rm lg Z_{El}$  \\  \hline \hline
Mg   & -3.1                 & 7.64   &  -1.57                       &  -4.67       & -0.38                     &  -3.48       \\ 
Fe   & -2.84                & 7.54    & -1.63                       & -4.47        & -0.48                     & -3.32       \\ 
Si   & -3.172                & 7.51   & -1.14                        & -4.31        & -0.18                     & -3.35       \\ 
Cr   & -4.75                 & 5.64   & -1.68                        & -6.43        & -0.57                     & -5.32       \\
Ca   & -4.192                &  6.34  & -1.34                        & -5.53        &  -0.32                    & -4.51       \\
\hline \hline
\end{tabular}
\label{tab:mfZ}
\end{table*}

There are several other studies in the literature, in which abundances of chemical elements in GCs were
determined using the integrated-light spectra. 
Abundances of several chemical elements were determined by \citet{larsen17} using the integrated-light 
spectra of seven~GCs. However, only two of them have high metallicity.
Likewise, in \citet{sharina20}, all GCs considered are of the low metallicity only.

\citet{Conroy18} measured abundances of several chemical 
elements using the medium-resolution integrated-light spectra of 41 GCs from \citet{Schiavon05} and their original method. 
We cannot use their results, because the distribution of $\rm \feh$ determined by \citet{Conroy18} do not show a 
separation into low- and high-metallicity components, while the library of \citet{Schiavon05} contains GCs with the 
\feh\ values higher and lower than -1~dex. The mean value  $\rm \langle \feh \rangle=-1.35$ \citep{Conroy18}
is higher by $\sim 0.3$ dex for low-metallicity GCs  and $\rm \langle \feh \rangle=-0.68$
\citep{Conroy18} is lower by $\sim 0.2$ dex for high-metallicity GCs  than the corresponding value in \citet{d16}, \citet{carretta10}, 
and \citet{pritzl05}
(see, please, \citet{ASh18}). 

In the following, we will provide the analysis of nucleosynthesis using the data from \citet{colucci17}. 
The statistical analysis of chemical abundances in the low- and high-metallicity groups of GCs from the paper by 
\citet{colucci17} is given in Table~\ref{tab:colucci17}.

It should be noted that the mean values of $\rm \mgh$ for the low- and high-metallicity groups of GCs according to the data 
obtained by \citet{colucci17} are close to those obtained by \citet{w16} for CGCs but systematically lower 
than the estimates derived using the data from  \citet{d16}, \citet{carretta10}, and \citet{pritzl05} (see Table~2 in  \citet{ASh18}). 
The mean $\rm \feh$ estimates agree with 
the results of the aforementioned studies. Let us show that the systematic shifts in the magnesium abundances 
do not noticeably influence the estimates of the SN~CC number. 

According to the reasoning by \citet{ASh18}, $20\% - 50\%$ of the cloud mass can be enriched as a result of  
nucleosynthesis in the first generation of GCs. In this case, from $\rm 10\%$ to $25\%$ of the gas mass is transformed into stars. 
{\it The results of nucleosynthesis in SNe~Ia obtained in the  following analysis do not depend on what fraction of the mass 
is enriched, since the number of different types of SNe and the mass of the produced chemical element will be directly 
proportional to this fraction.} This can be seen from the analysis of nucleosynthesis.

We will demonstrate in detail the algorithm for calculating the mass of a chemical element using magnesium as an example.
Let us estimate the mass fraction of magnesium corresponding to the average values of \mgh\ for metal-rich 
and metal-poor groups of GCs and CGCs.

Let $\rm Z_{Mg}$ denote the mass fraction of magnesium in relation to hydrogen in the object under study,
$\rm Z_{Mg_{\sun}}$ is the same mass fraction in the Sun. $\rm \mgh$ can mean both the fraction by the number 
of atoms $\rm N_{Mg}\over N_{H}$ and by the mass, where the mass fraction is determined with the expression 
$\rm N_{Mg} \cdot m_{Mg}\over N_{H} \cdot m_{H}$. 
By definition, $\rm \mgh = lg Z_{Mg} - lg Z_{Mg_{\sun}} $.
It follows from here that $\rm Z_{Mg} = 10^{lg Z_{Mg_{\sun}} + \mgh}$.

The mass fraction of magnesium in the Sun $\rm Z_{Mg_{\sun}} = {{N_{Mg} \cdot m_{Mg}}\over {{N_{H} \cdot m_{H}}}} X$,
 where $\rm X=0.7381$ is the mass fraction of hydrogen in the Sun \citep{asplund09}. It follows from here that 
 $\rm lg Z_{Mg_{\sun}} = lg ({N_{Mg}\over N_{H}}) + lg ({m_{Mg}\over m_{H}}) + lg(X) = -3.1$.
 (We used the value determined in the paper by \citet{asplund09} for the Sun 
 $\rm A_{Mg} = lg ({N_{Mg}\over N_{H}}) + 12 = 7.64$).
Thus, if the magnesium abundance in GCs $\rm \mgh=-1.57$, then the mass fraction of magnesium 
in it $\rm Z_{Mg}=10^{-1.57-3.1} = 10^{-4.67}$. 
If the magnesium abundance in GCs $\rm \mgh=-0.38$, then the mass fraction of magnesium 
in it $\rm Z_{Mg}=10^{-0.38-3.1} = 10^{-3.48}$. These values are given in Table~4.
Therefore, the magnesium abundance in the metal-rich GC group is fifteen times higher 
than that in the metal-poor group.

Similar to the procedure described above for finding the mass fraction of magnesium in the gas, from which GCs have been formed, 
we will summarize how we calculate the average mass fraction of the chemical element $El$.
Let us denote the mass fraction of the  element in the studied object as $\rm Z_{El}$, and the mass fraction of 
the element in the Sun as $\rm  Z_{{El}_{\sun}}$. Therefore, by definition: $\rm  [El/H] = lg Z_{El} - lg Z_{El_{\sun}}$ and
$\rm Z_{El_{\sun}} = {{N_{El} \cdot m_{El}}\over {{N_{H} \cdot m_{H}}}} X$, where $\rm N$ denotes the number of atoms along 
the line of sight. 
Hence,  $\rm lg Z_{El_{\sun}} = lg ({N_{El}\over N_{H}}) + lg ({m_{El}\over m_{H}}) + lg(X)$.  
To determine $\rm lg Z_{El_{\sun}}$, we use the values defined in the paper by \citep{asplund09} for the Sun:  $\rm X=0.7381$ 
is the mass fraction of
hydrogen in the Sun, therefore, $\rm A_{El}=lg({N_{El}\over N_{H}})+12 $. Thus, the mass fraction  
$\rm Z_{El}=10^{[El/H] + lg Z_{El_{\sun}}}$ 
corresponds to the mean abundance  $\rm  \langle [El/H]\rangle$ of the chemical element $\rm El$. 

  Table~\ref{tab:mfZ} shows the results of calculations of the mass fractions $\rm lg Z_{El}$ of the chemical elements 
Mg, Fe, Si, Cr, and Ca in GCs together with the values $\rm lg Z_{El_{\sun}}$ and $\rm A_{El}$.

In the next sections, we will calculate the number of SNe~CC and SNe~Ia that have exploded during the formation 
of the first generation of GCs from the 
known masses of the synthesized Mg and Fe. Then, using the determined number of SNe, we will calculate
the mass of Si, Cr, and Ca synthesized in the explosion of a single SN~Ia.
Each time we will consider the situation that $\rm 20\%$ of the cloud mass was enriched by SN explosions during 
the formation of low-metallicity GCs. 
 It will be shown below in the paragraphs 3.3--3.5 that the predictions of supernova nucleosynthesis do not 
 depend on the fraction of the enriched cloud mass.
The results of the analysis of the Si, Cr, and Ca production in GCs are presented in Tables~\ref{tab:PTD} 
and \ref{tab:W7}. These tables are organized as follows. The considered isotopes are indicated in the first row. 
The first column indicates 
what kind of data are considered: observed (values obtained in this paper) or theoretical (the values obtained 
using the supernova nucleosynthesis models). 
In the case of the observed data, the number of SNe is given, for which the mass of the synthesized chemical 
elements will be calculated. 
If the theoretical data are considered, the reference and the name of the model used are given.


\subsection{Number of SNe~CC responsible for the magnesium production in circumgalactic clouds}

If $\rm 20\%$ of the cloud mass was enriched by SN explosions during the formation of low-metallicity GCs, the cloud acquired 
the Mg mass $\rm M_{Mg}=0.2\cdot 10^{9}\cdot(10^{-3.48}-10^{-4.67}) \approx 10^{4.8} M_{\sun}$. 
 (If $\rm 50\%$ of the cloud mass was enriched, then the acquired $\rm Mg$ mass increased 
 2.5 times: $\rm M_{Mg}=0.5\cdot 10^{9}\cdot(10^{-3.48}-10^{-4.67}) \approx 2.5\cdot 10^{4.8} M_{\sun}$.)
Since SNe~Ia produce the negligible Mg 
mass (see the reasoning by \citet{ASh18}), then one could expect the number of SNe~CC: $\rm {10^{4.8}\over0.1}
= 10^{5.8}$. 
 (If $\rm 50\%$ of the cloud mass was enriched, then the number of SNe~CC increased 2.5 times:
$\rm {2.5\cdot 10^{4.8}\over0.1}= 2.5\cdot 10^{5.8}$.)
The first row in the observation section of Table~\ref{tab:PTD} shows the amount of SNe~CC determined based 
on the analysis of the Mg mass acquired by the cloud as a result of nucleosynthesis in the first generation of GCs
 under the assumption that $\rm 20\% $ of the cloud mass was enriched.


\subsection{Number of SNe~Ia responsible for the iron production in circumgalactic clouds}
If $\rm 20\%$ of the cloud mass was enriched by SN explosions during the formation of low-metallicity GCs,
the cloud acquired the  mass: $\rm M_{Fe}=0.2\cdot 10^{9}\cdot(10^{-3.32}-10^{-4.47}) \approx 10^{4.95} M_{\sun}$.
 (If $\rm 50\%$ of the cloud mass was enriched, then the number of SNe~Ia increased 2.5 times: 
$\rm M_{Fe}=0.5\cdot 10^{9}\cdot(10^{-3.32}-10^{-4.47}) \approx 2.5\cdot 10^{4.95} M_{\sun}$.) 

Since SNe~CC  produce the negligible Fe mass (about 6\%; see, please, the reasoning by \citet{ASh18}), 
then one could expect the number of SNe~Ia  depending on the iron mass ejected during the explosion of one SN~Ia:
$\rm {10^{4.95}\over0.23} = 10^{5.6}$  or ${10^{4.95}\over0.67} = 10^{5.1}$.
(If $\rm 50\%$ of the cloud mass was enriched, then the number of SNe~Ia will increase 2.5 times: 
$\rm 2.5\cdot {10^{4.95}\over 0.23} = 2.5\cdot 10^{5.6}$ or $ 2.5\cdot {10^{4.95}\over0.67} =2.5\cdot 10^{5.1}$). 
The value $\rm 0.23$ was obtained for the first time in the paper by \citet{AM13}
 employing the developed theory of the iron synthesis in the galactic disc, which was able to explain 
 the subtle features in its distribution.

 This value agrees with the conclusions drawn by \citet{Childress15} and is the mean value for pure deflagration models 
$\rm 050-1-c3-1P$ and $\rm 100-1-c3-1P$, in which carbon ignites at a lower central stellar density of the stellar remnant 
\citep{LeungNomoto18}. 
This situation is possible in the single degenerate scenario at a high accretion rate onto the carbon-oxygen WD that 
leads to a rapid temperature increase 
in the centre of a WD until it reaches the Chandrasekhar limit.
 An iron mass of $\rm 0.67 M_{\sun}$, first proposed by \citet{Nomoto84}, was obtained with the W7 and WDD2 models and 
 reconsidered for the updated 
 nuclear reaction network by \citet{LeungNomoto18}. 

It is important to note that due to the insufficient completeness of the samples of GCs, the obtained average values 
can be biased from the values 
that would be obtained for more complete samples. 
But, since there are no more complete samples, we will continue the study using the  available material.

\textit{Using the determined number of SNe of different types on the basis of the magnesium and iron abundances,
we will analyse the enrichment of the cloud with other chemical elements: silicon, chromium, and calcium.}


\subsection{Analysis of silicon abundances in GCs}
\label{sec:Si}
If $\rm 20\%$ of the cloud mass was enriched by SN explosions during the formation of low-metallicity GCs,
 the cloud has acquired the mass $\rm M_{Si}=0.2\cdot 10^{9}\cdot(10^{-3.35}-10^{-4.31}) \approx 10^{4.9} M_{\sun}$.
 Accordingly, if $5\rm 0\%$ of the cloud mass was enriched by SN explosions, the cloud acquired the mass: 
 $\rm M_{Si}\approx 2.5\cdot 10^{4.9} M_{\sun}$.

We will consider the mass of $\rm ^{28}$Si produced during a SN~CC explosion (Tables~\ref{tab:PTD} and \ref{tab:W7}) 
and the number of SNe~CC and pSNe~Ia as known.

Hence, it is obvious that the mass of silicon produced during the pSN~Ia explosion does not depend on what 
fraction of the cloud is enriched. 
For example, if $50\% $ of the cloud mass was enriched by SN explosions, $\rm  2.5 \cdot 10^{5.8} $ SNe~CC will produce 
$ \rm 2.5 \cdot 10^{5.8} \cdot 0.1 M_{\sun}$ of silicon, then the remaining mass of silicon falling onto pSNe~Ia  
$ \rm M_{Si} \approx 2.5 \cdot 10^{4.9} -2.5 \cdot 10^{5.8} \cdot 0.1(M_{\sun})$. This is $2.5$ times larger 
than the mass produced 
under the assumption that $ \rm 20\%$ of the cloud is enriched. However, this number is divided by the number of the pSN~Ia 
supernovae that is 
also 2.5 times greater: $\rm  2.5 \cdot 10^{5.6} $. Thus, the coefficient of 2.5 is reduced. Therefore, in Tables 5 and 6, 
all the average masses of chemical elements are given under the assumption that the $ 20\% $ of the cloud has become 
enriched, because the conclusions about nucleosynthesis in pSNe~Ia remain valid 
for any arbitrary fraction of the enriched part of the cloud.
Furthermore, the analysis of the nucleosynthesis of the remaining chemical elements will be carried out under 
the assumption that $ 20\% $ 
of the cloud has been enriched.
Tables~\ref{tab:PTD} and \ref{tab:W7}  present comparison between the masses of $\rm ^{28}$Si synthesized in a SN burst and 
calculated from the analysis of the corresponding elemental abundances in GCs from the paper by \citet{colucci17} with the 
results of theoretical studies of nucleosynthesis yields.
The second row of the observation section in Table~\ref{tab:PTD} shows the amount of SNe~Ia determined for the mass of iron 
$\rm M_{Fe} = 0.23 M_{\sun}$ synthesized during the explosion of a single SN.
 This mass of iron, as was already mentioned, is the average value for two-dimensional pure deflagration models $\rm  050-1-c3-1P$ 
 (it produces $\rm M_ {Fe} = 0.21 M_{\sun}$) and $ 100-1- c3-1P$ (it produces $\rm M_ {Fe} = 0.265 M_{\sun}$) \citep{LeungNomoto18}. 
Since we have determined that SNe~CC produce $\rm 10 ^ {4.8} M_{\sun}$ of $ Si $, SNe~Ia produce 
$ \rm M_ {Si} = 10 ^ {4.9} -10 ^ {4.8} = 10 ^ {4.2} M_{\sun} $. The mass of Si produced during one SN 
explosion with this amount of SNe~Ia $\rm 4.1 \cdot 10 ^{- 2}M_{\sun} $ is shown in the second row of the observation section 
of Table~\ref{tab:PTD}. Four rows of the theoretical data section of Table~\ref{tab:PTD} list the versions of the two-dimensional
pure deflagration models for different values of the central density of a WD by \citet{LeungNomoto18} and the corresponding 
mass of $\rm ^{28}$Si synthesized during the explosion of a WD. 

The data in the first column of the observation section in Table~\ref{tab:W7} indicate the amount of SNe~Ia determined for 
the mass of iron $\rm M_ {Fe} = 0.67 M _ {\ sun} $ synthesized during the explosion of a single SN.
This mass of iron corresponds to the one-dimensional Chandrasekhar mass deflagration model W7 \citep{Nomoto84} and the model with 
deflagration-detonation transition WDD taking into account the updated nuclear reaction network \citep{LeungNomoto18, Mori18}. 
The first column of the observation section of Table~\ref{tab:W7} shows the mass of $\rm ^{28}$Si in the explosion of one SN~Ia 
which we obtained based on the analysis of GCs with this number of SNe~Ia. While the four rows of the 
first column in the theoretical data
section of Table~\ref{tab:W7} show the mass of $\rm ^{28}$Si synthesized during the explosion of a WD corresponding to 
the indicated models \citep{LeungNomoto18, Mori18}.
\begin{table}
 \centering
\caption{Comparison between the masses of silicon $\rm ^{28}$Si, chromium $\rm ^{52}$Cr, and calcium 
$\rm ^{40}$Ca synthesized in a SN burst and calculated from the analysis of the corresponding elemental abundances in GCs 
\citep{colucci17} 
and the results of theoretical studies of nucleosynthesis yields of two-dimensional pure turbulent deflagration (PTD) 
models for various values of the central density of a WD by \citet{LeungNomoto18} (LN18) }

\begin{tabular}{lccc} \hline \hline
Isotope/              &           $\rm^{28}$Si  & $\rm^{52}$Cr            & $\rm^{40}$Ca \\
Source & \multicolumn{3}{c}{$\rm <Mass> (M_{\sun})$ }          \\ \hline \hline
 Observed data  &                      &           &                                    \\ 
$\rm N_{SN~CC}=10^{5.8}$  &   0.1                & $\rm 1.20 \cdot 10^{-3}$ & $\rm 5.6 \cdot 10^{-3}$ \\ 
 $\rm N_{SN~Ia}=10^{5.6}$  & $\rm 4.1 \cdot 10^{-2}$ & $\rm 2.5 \cdot 10^{-4}$  & $\rm 5.0 \cdot 10^{-3}$ \\ \hline 
Theoretical data    &                      &           &                                 \\ 
{\underline{PTD LN18 models}}      &                      &           & \\
050-1-c3-1P           & $\rm  3.57 \cdot 10^{-2}$& $\rm  7.53 \cdot 10^{-4}$& $\rm  1.71 \cdot 10^{-3}$ \\
  100-1-c3-1P         & $\rm  3.82 \cdot 10^{-2}$& $\rm  1.11 \cdot 10^{-3}$& $\rm  1.82 \cdot 10^{-3}$  \\
  300-1-c3-1P         & $\rm  3.46 \cdot 10^{-2}$& $\rm  4.46 \cdot 10^{-3}$& $\rm  1.50 \cdot 10^{-3}$ \\
  500-1-c3-1P         & $\rm  3.23 \cdot 10^{-2}$& $\rm  1.20 \cdot 10^{-2}$& $\rm  1.50 \cdot 10^{-3}$ \\ \hline\hline
 \end{tabular}
\label{tab:PTD}
\end{table} 

\begin{table}
 \centering
\caption{Comparison between the masses of silicon $\rm  ^{28} Si $, chromium $\rm^{52}$Cr, 
and calcium $\rm^{40}$Ca synthesized 
in a SN burst and calculated from the analysis of the corresponding elemental abundances 
in GCs \citep{colucci17} and the results of theoretical studies of
nucleosynthesis yields of the one-dimensional Chandrasekhar mass deflagration model W7 
(for $Z=0.1Z_sun$ ) and the model with 
deflagration-detonation transition WDD  by \citet{LeungNomoto18} and \citet{Mori18} }

\begin{tabular}{lccc} \hline \hline
Isotope/              &           $\rm^{28}$Si  & $\rm^{52}$Cr            & $\rm^{40}$Ca \\
Source & \multicolumn{3}{c}{$<Mass> (M_{\sun})$ }          \\ \hline \hline
 Observed data  &                      &           &                                    \\ 
$\rm N_{SN~Ia}=10^{5.1}$  &  $\rm 1.3 \cdot 10^{-1}$ & $\rm 8.0 \cdot 10^{-4}$  & $\rm 1.6 \cdot 10^{-2}$ \\ \hline 
Theoretical data    &                      &           &                                 \\ 
{\underline{LN18}}   &                     &           & \\
  W7                  & $ \rm 1.34 \cdot 10^{-1}$& $\rm  8.31 \cdot 10^{-3}$ & $\rm  1.42 \cdot 10^{-2}$ \\
WDD2                & $\rm  2.28 \cdot 10^{-1}$& $\rm  1.54 \cdot 10^{-2}$& $\rm  2.47 \cdot 10^{-2}$ \\ 
{\underline{Mori+2018}} &                      &           & \\
  W7                  &$ \rm 1.7 \cdot 10^{-1}$  & $\rm  7.2 \cdot 10^{-3}$ & $\rm  1.13 \cdot 10^{-2}$  \\
  WDD2                & $\rm  2.3 \cdot 10^{-1}$ & $\rm  1.35 \cdot 10^{-2}$& $\rm  2.48 \cdot 10^{-2}$  \\ \hline \hline
 \end{tabular}
\label{tab:W7}
\end{table} 

We can draw the following conclusion based on the comparison of the data in Tables~\ref{tab:PTD} and \ref{tab:W7}.
If in the analysis of iron enrichment we take the mass corresponding to its synthesis in pure turbulent deflagration 
models, that is, $\rm 0.23 M_{sun}$,
then we get $\rm 10^{5.6}$ SNe~Ia. This amount of SNe will lead to the mass of $\rm^{28}$Si ejected during the explosion of 
a single SN $\rm 4.1 \cdot 10^{-2} M_{\sun}$ which corresponds to the theoretically predicted value for the same pure turbulent 
deflagration models. If we use the $\rm^{56}Fe$ mass corresponding to the W7 or WDD2 models, then the $\rm^{28}$Si  mass corresponds 
to the W7 model. As can be seen from Table~9 of \citet{LeungNomoto18}, the W7 and WDD2 models provide close mass estimates only in the case 
of the Fe production. The WDD2 model provides the mass $1.5-2$ times higher than that from the W7 model for other chemical elements.

To summarise, the $\rm^{28}Si$ masses synthesized during a single SN~Ia  explosion are fully consistent with 
the theoretical nucleosynthesis 
calculations of the pure turbulent deflagration models of nucleosynthesis, both under the assumption of the reduced  
central density and for the W7 model.


\subsection{Analysis of chromium abundances in GCs}
\label{sec:Cr}
If $20\%$ of the cloud mass was enriched by SN explosions during the formation of low-metallicity GCs,
then the cloud acquired the mass   $\rm M_{Cr}=0.2\cdot 10^{9}\cdot(10^{-5.32}-10^{-6.43}) \approx 10^{2.91} M_{\sun}$.

Further, we will consider the mass of $\rm^{52}$Cr  produced during a SN~CC explosion (Tables~\ref{tab:PTD} 
and \ref{tab:W7}) and the number of SNe~CC and pSNe~Ia as known. Our reasoning will be analogous to that in the case 
of $\rm^{28}$Si considered
in the previous section (see also Tables~\ref{tab:PTD} and \ref{tab:W7}). We find that SNe~CC produce $\rm 10^{2.87} M_{\sun}$
of $\rm^{52}$Cr. Therefore, SNe~Ia produce $\rm M_{Cr}=10^{2.91}-10^{2.87}= 10^{1.9} M_{\sun}$. 
Depending on the number of  SNe~Ia, the 
chromium mass produced during a single SN explosion can be equal to $\rm2.5\cdot10^{-4} M_{\sun}$ or $\rm8\cdot10^{-4} M_{\sun}$.

The mean chromium mass produced by a single SN~Ia basing on the assumption that $\rm 10^{5.6}$  SNe~Ia exploded in the clouds 
is equal to $\rm 2.0 \cdot 10^{-4}M_{\sun}$. It is $3.5$ times lower than the minimum mass predicted by the theoretical models, 
as it follows 
from the pure turbulent deflagration model $\rm 050-1-c3-1P$ from \citet{LeungNomoto18}. The mean chromium mass during $\rm 10^{5.1}$ SN~Ia 
explosions is an order of magnitude lower than the value predicted with W7 and WDD2.

 To summarise, additional studies are necessary in the case of chromium. So, due to a small amount of GCs in the \citet{colucci17} 
 sample, the average values of magnesium and iron in the high- and low-metallicity groups of the considered chemical 
 elements can be biased, one can assume that the amounts of SNe~CC and SNe~Ia may differ from the calculated ones. However, in
 this case, first, the simultaneous coincidence of theory and observations for iron and silicon production is violated, and
 second, in order to harmonize the theory and observations for calcium (see below), the amount of SNe~Ia should not
 be $3.5$ time reduced, 
 but vice versa, three times increased. If we assume that the difference between the average values of  
 the chromium abundance of the low- and high-metallicity subgroups is by $0.1$~dex greater than it follows from the 
 \citet{colucci17} data; then we can reconcile the result of the SN~Ia nucleosynthesis with the predictions
of pure turbulent deflagration models for lower central density values. The value of $0.1$~dex sufficient for the agreement is 
several times smaller than the standard deviation $\rm \sigma(\crh)$ as can be seen from 
Table~\ref{tab:colucci17}.


\subsection{Analysis of calcium abundances in GCs}
\label{sec:Ca}
If $\rm 20\%$ of the cloud mass was enriched by SN explosions during the formation of low-metallicity GCs, 
the cloud acquired the mass of $\rm^{40}$Ca: $\rm M_{Ca}=0.2\cdot 10^{9}\cdot(10^{-4.51}-10^{-5.53}) \approx 10^{3.75} M_{\sun}$.

Further, we will consider the $\rm^{40}$Ca mass produced during a SN~CC explosion
(Tables~\ref{tab:PTD} and \ref{tab:W7}) and the number of SNe~CC and pSNe~Ia as known.
Our reasoning will be analogous to that in the case of $\rm^{28}$Si considered in Sec.~\ref{sec:Si} 
(see also Tables~\ref{tab:PTD} and \ref{tab:W7}).
 We obtain that SNe~CC produce $\rm 10^{3.55} M_{\sun}$ of $\rm^{40}$Ca. Therefore, SNe~Ia produce 
 $\rm M_{Ca}=10^{3.75}-10^{3.55}= 10^{3.3} M_{\sun}$.
 Depending on the number of SNe~Ia, the calcium mass produced during a single SN explosion can be equal to
 $\rm 5.0\cdot 10^{-3} M_{\sun}$ or $\rm 1.6\cdot 10^{-2} M_{\sun}$.

A good agreement can be reached for the calcium mass with the W7 results (Table~\ref{tab:W7}). 
The mass of calcium produced by  
$\rm 10^{5.6}$ SNe~Ia is equal to $\rm 5.0 \cdot 10^{-3} M_{\sun}$ that is $3$ times larger than the value predicted 
in the two-dimensional 
pure turbulent deflagration models. 

However, one can notice the following trend: the increase in the synthesized calcium mass and at the same time reducing 
the synthesized chromium mass can be achieved with the decrease in the burning WD density (see, please, Table~5).


\subsection{Chemical abundances according to  the \citet{pritzl05} data}
\label{sec:pritzl}
In order to check the dependence of the estimates obtained in Secs.~\ref{sec:Si}--\ref{sec:Ca} on the results of observations, 
we will take similar reasoning for the data by \citet{pritzl05}. The number of GCs in this study is larger, but we cannot 
completely rely on them for the following reason. 

Using the same algorithm as that described in Sec.~\ref{sec:analysis}, we conclude that 
the cloud has acquired the magnesium mass $\rm M_{Mg}= 10^{4.93} M_{\sun}$.
Therefore, $\rm 10^{5.93}$ SNe~CC have exploded.  We can conclude that the cloud has acquired the iron mass 
$\rm M_{Fe}= 10^{4.88} M_{\sun}$. 
If the mass of the synthesized iron in a single SN explosion is equal to $\rm 0.23 M_{\sun}$, then $\rm 10^{5.52}$ 
SNe~Ia have exploded.
If the  mass of the synthesized iron in a single SN explosion is equal to $\rm 0.67 M_{\sun}$, then $\rm 10^{5.05}$ 
SNe~Ia have exploded. 

Then, using the analysis analogous to that described in Sec.~\ref{sec:Si}, we conclude, that the cloud has acquired 
$\rm M_{Si}= 10^{4.88} M_{\sun}$. 
One can see that a smaller amount of Mg than Si was produced.  Considering that the average masses of magnesium and silicon
during a single SN~CC explosion are equal to $\rm 0.1 M_{\sun}$,
the silicon production cannot be explained, because there is no substance left for the contribution from a SN~Ia. 
For example, if we increase the average value of the magnesium in the low-metallicity subgroup by 0.4 dex, 
we will reduce the mass of the synthesized magnesium and there will be no conflict with the mass of the synthesized silicon.

Let us analyse the possible reasons for this situation. 
 \citet{pritzl05} considered mostly remote low-metallicity GCs. Additionally, as was mentioned in Sec.~3.2, \citet{pritzl05} 
 considered the abundances of bright red giants in the clusters which cannot be 
representative for all stars in GCs at all evolutionary stages. High- and low-metallicity  subgroups fall into different 
subsystems of the Galaxy: several disc and bulge GCs are in the high-metallicity subgroup (see Table~\ref{tab:pritzl}), 
halo GCs are in the low-metallicity  subgroup. 

\begin{table*}
 \centering
\caption{Abundances of chemical elements in GCs according to \citet{pritzl05}}
\begin{tabular}{@{}l|lcc|lcc@{}}
\hline \hline
Chemical  & \multicolumn{3}{c}{\underline{Low-metallicity subgroup ($\rm [X/H] <-1$)} } & \multicolumn{3}{c}{\underline{High-metallicity subgroup ($\rm [X/H] \ge-1$)}} \\ 
 element  &  $\rm \langle [X/H] \rangle$ &  $\rm \sigma([X/H])$ & $ N_{GC}$ &  $\rm \langle [X/H] \rangle$ &  $\rm \sigma([X/H])$ & $ N_{GC}$  \\ \hline\hline
         & $(\rm \mgh<-1.0)$  &  &  &      $(\rm \mgh \ge -1.0)$ & & \\ 
$\rm^{24}$Mg & -1.38 & 0.42 & 33 & -0.24 & 0.23 & 6 \\ \hline
        & $\rm (\sih<-0.7)$  &  &  &     $\rm (\sih \ge -0.7)$  & &  \\ 
$\rm^{28}$Si     &-1.34 & 0.25 & 26 & -0.22 & 0.27 & 6 \\ \hline
        & $\rm (\cah<-0.8)$  &  &  &            $\rm (\cah \ge -0.8)$  & &  \\ 
$\rm^{40}$Ca  & -1.42 & 0.28 & 27 & -0.29 & 0.27 & 6 \\ \hline
        & $\rm (\feh<-1.0)$  &  &  &            $\rm (\feh \ge -1.0)$  & &  \\ 
$\rm^{56}Fe$  & -1.73 & 0.37 & 32 & -0.55 & 0.29 & 6 \\ \hline\hline
\end{tabular}
\label{tab:pritzl}
\end{table*}

The situation appeared to be better for calcium (Table~\ref{tab:7} and \ref{tab:Ca}). 
Then, using the analysis analogous to that described in 
Sec.~\ref{sec:Ca}, we  conclude, that the cloud has acquired $\rm M_{Ca}= 10^{3.8} M_{\sun}$, 
and SNe~CC produced the mass  of Ca $\rm 10^{3.7} M_{\sun}$.
It follows from here that SNe~Ia  produce $\rm M_{Ca}=10^{3.8}-10^{3.7}=10^{3.11}M_{\sun}$. 
Depending on the number of SNe~Ia,
the calcium mass ejected in a single SN explosion can be equal to $\rm 3.9\cdot10^{-3} M_{\sun}$ or $\rm 1.15\cdot10^{-2} M_{\sun}$. 
The resulting masses coincide with the theoretical ones only by an order of magnitude. 
One can achieve either full agreement
with the results of the two-dimensional pure turbulent deflagration (PTD) models, or with the 
W7 results within error of the average value determination.

\begin{table}
 \centering
\caption{Comparison between the mass of calcium $\rm^{40}$Ca produced during a SN 
burst and calculated from the analysis of the corresponding elemental abundances 
in GCs \citep{pritzl05} and the results of theoretical studies of nucleosynthesis 
yields of the two-dimensional pure turbulent deflagration 
(PTD) models for various values of the central density of a WD by \citet{LeungNomoto18} (LN18)}
\begin{tabular}{lc} \hline \hline
Isotope/              &           $\rm^{40}$Ca  \\
Source & $\rm <Mass> (M_{\sun})$                 \\ \hline \hline
 Observed data  &                          \\ 
$\rm  N_{SN~CC}=10^{5.8}$  & $\rm 5.8 \cdot 10^{-3}$  \\ 
 $\rm  N_{SN~Ia}=10^{5.6}$  & $\rm 3.9 \cdot 10^{-3}$ \\ \hline 
Theoretical data    &                         \\ 
{\underline{PTD LN18 models}}                          \\
050-1-c3-1P           & $\rm  1.71 \cdot 10^{-3}$ \\
  100-1-c3-1P         & $\rm  1.82 \cdot 10^{-3}$  \\
  300-1-c3-1P         & $\rm  1.50 \cdot 10^{-3}$ \\
  500-1-c3-1P         & $\rm  1.50 \cdot 10^{-3}$ \\ \hline\hline
 \end{tabular}
\label{tab:7}
\end{table} 
\begin{table}
 \centering
\caption{Comparison between the calcium $\rm^{40}$Ca masses synthesized in a SN burst and calculated from the analysis of 
the corresponding elemental abundances in GCs \citep{pritzl05} and the results of theoretical studies of nucleosynthesis yields 
of the one-dimensional Chandrasekhar mass deflagration model W7 and the model with
deflagration-detonation transition WDD  by \citet{LeungNomoto18} and \citet{Mori18}}
\begin{tabular}{lc} \hline \hline 
Isotope/              &           $\rm^{40}$Ca    \\
Source &       $\rm <Mass> (M_{\sun})$             \\ \hline \hline
 Observed data    &                        \\ 
$\rm  N_{SN~Ia}=10^{5.1}$   &  $\rm 1.15 \cdot 10^{-2}$  \\ \hline 
Theoretical data       &                        \\ 
{\underline{LN18}}     &                        \\
  W7                   & $\rm  1.42 \cdot 10^{-2}$  \\
WDD2                   & $\rm  2.5 \cdot 10^{-2}$  \\ 
{\underline{Mori+2018}}&                        \\
  W7                   &$\rm  1.13 \cdot 10^{-2}$  \\
  WDD2                 & $\rm  2.48 \cdot 10^{-2}$ \\ \hline \hline
 \end{tabular}
\label{tab:Ca}
\end{table} 


\section[]{Conclusions}

In this paper, we have proposed a method for determining the properties of type Ia supernovae from 
short-lived precursors -- Prompt SN~Ia. This method is based on the assumption that this very subtype 
of type Ia supernovae is responsible for the enrichment of the high-metallicity subgroup of globular clusters 
and circumgalactic clouds  with the iron peak elements.
 We believe that GCs are the most suitable laboratories to study the Prompt SN~Ia nucleosynthesis. 
First, the average age of the metal-rich and metal-poor GCs are about the same \citep{ch12, vandb13}.
Second, the density of stars in globular clusters is high. Hence, it can be expected that 
the interaction of stars in them will occur more often than in other stellar groups. The interaction 
of a white dwarf with its companion`s matter is an essential condition for all SN~Ia explosion models 
existing in the literature (e.g.,\citet{LeungNomoto18} and references therein). As was shown by \citet{ASh18}, 
the occurrence of Prompt SN~Ia bursts in GCs is $\rm 1.25 - 2.5$ times higher than that in the disc.
In addition, the evolution of multiple stars allows, at least for one of them, a carbon-oxygen core to form 
faster with a mass of about $\rm 1.4 M_{\sun}$ required by theory and capable of a subsequent explosion in 
comparison with a single star \citep{anguiano20}. This condition is important because a young progenitor 
population produces SNe~Ia on timescales of $\rm 100-330$~Myr (e.g., \citet{Aubourg08, MaozBadenes10} 
and references therein). 


 The accuracy of the method depends on 
the number of globular clusters, in which one method has determined as many such chemical elements as possible, 
which are produced only during supernova explosions.

The analysis of the GC  chemical enrichment, together with the characteristics of the parent clouds, allowed us to draw 
conclusions about the nucleosynthesis in the SNe~Ia concentrated in the star-formation regions, namely Prompt SNe~Ia.
There is no generally accepted precursor model for this SN~Ia subtype. It turns out that the nucleosynthesis in 
globular clusters  agrees with the pure turbulent deflagration models, both two-dimensional with a central density 
of $\rm 0.5-1~g/cm^3$ and one-dimensional W7. 
Based on the assumptions underlying the explosive 
nucleosynthesis  models, it can be concluded that the progenitors of Prompt SNe~Ia are best described 
by the Single Degenerate
scenario, in which a degenerate carbon-oxygen stellar remnant accretes the matter from the companion -- a Main 
Sequence or red giant star. 

 For more accurate conclusions, in order to choose between the two pure turbulent deflagration models, 
 homogeneous data are necessary on various 
 chemical elemental abundances in high- and low-metallicity groups of  GCs.
 However, it is important to emphasize that estimates of the production of Mg, Si, Ca, Cr, and Fe obtained 
 from analysis of the chemical composition of GCs and theoretical predictions of nucleosynthesis models in supernovae 
 are consistent. This is an independent confirmation of the correct approach to modelling nucleosynthesis.
   
We find it important to note that the model of Single Degenerate progenitors meets another difficulty: 
a hydrogen mass of about $\rm 0.01 M_{\sun}$
detected using the $\rm H\alpha$ emission line in the spectra of several  SNe~Ia is much lower 
than that expected for the Main Sequence+WD or Red Giant+WD progenitors.

The result allows us to understand in a new way the absence of  changes of the \afe\ values with 
time during the evolution of a GC. 
Only SNe~Ia producing a small amount of iron are responsible for the elemental enrichment in GCs.
 
 In the process of solving the main issue, several phenomena related to our study were indicated, the 
further study of which may add important refinements to the theory of the formation and evolution of the galaxy.
 
These include: rarely seen CGCs and GCs with $\rm \langle [X/H] \rangle < -2.3$, the lack of very 
high-metallicity GCs and CGCs with $\rm [X/H] > -0.3$. The paper discusses the conditions for the formation 
of star clusters, the masses of which are comparable to the GC initial mass $\rm 10^6-10^7 M_{\sun}$ in near-galactic clouds.
 
\section*{Acknowledgments}
We thank the anonymous referee for comments that helped to improve the paper. 

\section*{Data Availability}
The data underlying this article are available in the article.

\bsp

\label{lastpage}

\end{document}